# Phishing Attacks and Websites Classification Using Machine Learning and Multiple Datasets (A Comparative Analysis)

Sohail Ahmed Khan[1], Wasiq Khan[2], Abir Hussain[3]

[1] The University of Sheffield, Sheffield S102TN, UK
`sohailahmedkhan173@gmail.com`
[2] Liverpool John Moores University, Liverpool L3 5UG, UK
`w.khan@ljmu.ac.uk`; `A.Hussain@ljmu.ac.uk`

**Abstract.** Phishing attacks are the most common type of cyber-attacks used to obtain sensitive information and have been affecting individuals as well as organizations across the globe. Various techniques have been proposed to identify the phishing attacks specifically, deployment of machine intelligence in recent years. However, the deployed algorithms and discriminating factors are very diverse in existing works. In this study, we present a comprehensive analysis of various machine learning algorithms to evaluate their performances over multiple datasets. We further investigate the most significant features within multiple datasets and compare the classification performance with the reduced dimensional datasets. The statistical results indicate that random forest and artificial neural network outperform other classification algorithms, achieving over 97% accuracy using the identified features.

## 1      Introduction

Phishing in general, is a fraud in which a target (e.g., person) or multiple targets are contacted by email, telephone or text message by a fraudster or cybercriminals [1]. These cybercriminals pose as a legitimate and reputable entity or a person and try to convince individuals into giving up their sensitive data such as passwords, identity information, bank or credit card details. The provided information is then used to gain access to important accounts or services and can result in identity theft and financial loss. Phishing is popular among fraudsters due to its simplicity to trick somebody into clicking a malicious link and trying to break through a computer's defence systems or to bypass modern authentication systems.

Variety of attributes have been used to identify a phished webpage such as use of IP address in the URL, abnormal URL (special symbols in the URL) and many more [2]. However, a naive computer user can easily be tricked into considering a fake webpage as a legitimate webpage. Various techniques have been employed to deal with phishing attacks and distinguishing the phishing webpages automatically. For instance, blacklist-based detection technique keeps a list of websites' URLs that are categorized as phishing sites. If a web-page requested by a user exists in the formed list, the connection to the queried website is blocked [2].

The web-page feature based approach (i.e., visual features) [3] examines the abnormalities in web-pages, such as, the disagreement between a website's identity and its structural features. Likewise, Machine Learning (ML) based approaches rely on classification algorithms such as Support Vector Machines (SVM) [5] and Decision Trees (DT) [4] to train a model that can later automatically classify the fraudulent websites at run-time without any human intervention. Most of the existing studies focuses single classifiers and/or a single dataset however, it would be helpful to investigate different classifiers while using multiple dataset with variety of attributes. Likewise, investigation of the most significant features within the multiple datasets might be of special interest.

This manuscript entails a comprehensive review of different ML algorithms for the phishing web-sites classification. Compared to existing research, we present a review study that performs the comprehensive analysis and comparison of different techniques for the classification of phishing websites. For the first time in this study, we used three different datasets to train, test and validate multiple classification algorithms including DT[4], SVM[5], Random Forest[22], Naïve Bayes (NB) [23], K Nearest Neighbours (KNN) [24] and Artificial Neural Networks (ANN) [17], to distinguish the phishing websites from legitimate websites. We further employ well-known Principal Component Analysis (PCA) [6] for dimensionality reduction and achieves approximately similar even or better classification performance as compared to using full features within the dataset. In addition, we investigated the level of significance for all attributes/features within the three datasets using the PCA based component loadings. Rest of the manuscript is organized as follows. Section 2 addresses the existing work in this domain while Section 3 comprises the proposed methodology. Section 4 presents the discussion and comparison of results achieved followed by Section 5 which concludes the study and presents directions for future work.

## 2 Literature Review

Phishing websites detection is a crucial step towards countering online fraud. Recent technological advancements have been made with the use of ML and data science methods in diverse application domains including aerospace [28], speech processing [29], healthcare technologies [30, 32], border security [31], object recognition [33], cyber-crime detection [27], smart city [35] and so on. Likewise, there have been many technological developments in the domain of cyber security specifically to automatically detect the phishing attacks, but there is still a room for a lots of improvements in this regard. Malicious attackers are coming up with new techniques, and phishing incidents are on a rise [34]. Several detection strategies have been devised in order to counter phishing attacks. For instance, in [7], authors used dataset [8] containing 30 different types of features with 11055 instances. Authors trained five different classification algorithms i.e. prism, NB, K*, RF, and ANN. They achieved best results 98.4% and 95.2% with RF, both with and without feature selection respectively. The study achieved some very good results, but as it uses single dataset only, the results are not reliable enough to be universal. Zhang et al [11] employed ANN to detect phishing emails. The dataset they used was comprised of approximately 8762 emails, out of which 4560 were phishing

emails and rest were non-phishing or legitimate. They trained a feedforward neural network using resilient propagation training and compared the performance with other ML models. They found that while maintaining highest recall, ANN had 95% accuracy [26], which makes ANNs excellent at distinguishing phishing emails whilst misclassifying a slight percentage of legitimate emails. The study focuses the use of ML to classify phishing emails and did not considers the phishing websites classification.

In [12], Mohammad et al employed a rule-based classification technique for detecting phishing websites. They made use of 4 different classification algorithms where the study indicated that using the feature reduction algorithm and classification based association together produced the optimistic performance. The study only relied on features which were occurring in high frequency, which can be misleading sometimes, as higher frequency does not always guarantee higher importance. Karnik et al [13] used the SVM in combination with cluster ensemble to classify phishing and malware website URLs. Training is performed through SVM using kernel functions (Linear, Radial, Polynomial and Sigmoid). With the proposed technique, the SVM model predicted correctly 95% of the times. The study only takes URL based and textual features in to consideration, and does not consider any other feature such as host based and content (iframes etc.) based features. Likewise, other ML algorithms could be used for the comparative analysis to achieve more reliable findings.

A meta-heuristic based nonlinear regression algorithm for feature selection and phishing website detection is introduced in [14]. For classification, non-linear regression based on harmony search technique and SVM are deployed. Phishing dataset from UCI's machine learning repository [8] is used which contains 11055 instances and 30 features. This study did achieve some interesting results however, relied on one dataset only which itself is very small containing only 11055 instances. In [15], Sahingoz et al., proposed a real time phishing detection system based on 7 different classification algorithms and Natural Language Processing (NLP) based features (i.e. Word Vectors, NLP based and Hybrid features). They found the RF algorithm based only on NLP features to be the best performer with an accuracy of 97.98%. For testing, authors used classification algorithms implemented in WEKA suite. A similar work is presented in [16] that proposes phishing website classification based on a hybrid model to overcome the problem posed by phishing websites. They used the dataset [8] from UCI's machine learning repository which comprises of 30 features and 11055 total instances. The system achieved 97.75% accuracy using DT (J48) and ensemble method. Similar to other works, only single dataset was employed to train and test the algorithms. No feature selection/dimensionality reduction mechanism was implemented.

A heuristic based phishing detection technique is proposed in [17] which uses dataset of 3,000 phishing site URLs and 3,000 legitimate site URLs. Authors employed several ML algorithms including, KNN, RF, NB, SVM, and DT. Study indicated the RF to be the highest performer in all three performance measurements with an accuracy of 98.23%. however, this work also considers URL based features and does not consider other features such as content and domain-based features. Also, the training and test sets are very small i.e. dataset comprises of only 6000 instances.

Aforementioned research studies demonstrate a considerable amount of work has already been done in phishing websites classification using different ML based techniques. Researchers have employed different techniques in order to predict phishing websites efficiently and with better accuracy. However, it would be helpful to analyze the ML algorithms' performances over the multiple datasets as well as over the reduced features from all dataset to investigate the impact of dimensionality reduction on the classification performances. This work therefore employs PCA for the attribute analysis and dimensionality reduction on the three datasets used in this study and compares the classifiers' performances with results achieved using non compressed feature sets.

## 3 Methodology

### 3.1 Datasets

We used three different datasets in this study to investigate the ML algorithms' performances as well as the attribute importance within the three datasets. Dataset1 [10] comprises of 48 different features obtained from 5000 different phishing webpages and 5000 different legitimate webpages. The webpages were downloaded during the time period between January to May 2015 and from May to June 2017 [10]. This dataset is labelled with binary labels e.g. 0 for Legitimate, and 1 for Phishing. So for example, if the classifier predicts label 1 for a given website, this means the website is Phishing and vice versa. Dataset 2 is obtained from University of California, Irvine's Machine Learning Repository [8]. This dataset contains 30 different features which uniquely identify phishing and legitimate websites. The target variable is binary, -1 for Phishing and 1 for legitimate. The dataset is populated from different sources, some are PhishTank archive, Google search engine, and MillerSmiles archive. This dataset contains mostly the same features as dataset 1 with some additional features. In total, it contains 11055 distinct website entries out of which 6157 are legitimate websites and 4898 are phishing websites. The dataset features are normalized and given values from -1 to 1, where -1 represents Phishing, 0 represents suspicious and 1 means legitimate.

Dataset 3 [9] is obtained from University of California, Irvine's Machine Learning Repository [9] and contains different features related to legitimate and phishing websites. This dataset contains data from 1353 different websites collected from different sources. The collected data contains 702 phishing URLs, 548 legitimate URLs, 103 suspicious URLs out of total 1353 records. This is a multi-class dataset, i.e. three different class labels where -1 means phishing, 0 means suspicious, 1 means legitimate. Suspicious represents means a webpage can be either phishing or legitimate.

### 3.2 Experimental Design

The experiments are designed by utilizing different ML and data analytics libraries including Scikit Learn [18], KERAS [19], Numpy [20] and Pandas [21]. Five ML algorithms namely, decision tree [4], SVM [5], random forest [22], NB [23], n nearest neighbours (KNN) [24] and artificial neural networks (ANN) [17] were employed along with the PCA [6] based feature importance measure as well as reduced dimensions. For the baseline experimental setup, recursive classification trials are conducted to compare the classifiers' performances for the model tuning and configurations such as kernel (e.g. radial, polynomial), cost, gamma, ntree, number of neurons in each layer, batch size, and time stamp. Standard 10-fold cross-validation [25] train/test trials were run by partitioning the entire dataset into training and testing proportions of 70% and 30%, respectively. It was ensured that the test data contains fair distribution for all classes. Following the baseline experimental results, the classifiers' parameters were set imperially to get the optimistic performance. Following experiments are designed with a consistent classifiers' configurations:

- Train and test the five ML algorithms over the individual datasets (i.e. Dataset 1, Dataset 2 and Dataset 3) using 10-fold CV to compare the performances.
- Train and test the five ML algorithms over the PCA based dimension reduced datasets (PCs covering 90% of variance distribution) using 10-fold CV to compare the performances.

Additional experiments are conducted to investigate the attribute/feature importance in each individual dataset that represent the most distinguishing attributes to classify the phishing websites. The performances of ML models are assessed using various gold standards including accuracy, specificity, precision, recall and F1-score. Algorithm 1 summarizes the experimental steps carried out to conduct the above-mentioned experiments (A, B).

---

**Algorithm 1**

Let $S$ be a set of attributes for the phishing website dataset
$S=\{RandomString, RandomString, NumUnderscore .....\}$,
Let $C_1, C_2$ and $C_3$ be the set of the website's classes
Let *Legitimate*, *Phishing*, and *Suspicious* are the websites classes where:
*Phishing* = $\{c_1: c_1 \in S$ & website type = -1$\}$
*Legitimate* = $\{c_2: c_2 \in S$ & flood severity = 1$\}$
*Suspicious* = $\{c_3: c_3 \in S$ & flood severity = 0$\}$
Let $S$ be the set of PCA components where:
$S = \{s: \forall s \in S, \exists th \Rightarrow PC(s) > th\}$
*Training* = $\{t \in S\}$ where Training is 60% of $S$
*Test* = $\{ts \in S$ & $ts \notin Training\}$ where Test is 20% of $S$
*Validation* = $\{vs \in S$ & $vs \notin Training$ & $ts \notin test\}$ where *Validation* is 20% of $S$
For every selected ML algorithm determined
$E[Accuracy_{C1,C2,C3}] = \{S: S \Rightarrow ML(Training, Validation, Test)\}$

---

### 3.3 Feature Importance and Dimensionality Reduction

One of the well-known dimensionality reduction technique is PCA [6] that have successfully been deployed in various application domains [33]. Major aim of the PCA is to transform a large dataset containing large number of features/variables to a lower dimension which still holds most of the information contained in the original high dimensional dataset. The interesting property of PCA is the attribute loadings that can also be used for the identification of attribute importance within the original dataset. We utilized PCA for the dimensionality reduction as well as calculation of feature importance score to investigate the most distinguishing features within all three Datasets we used in this study.

Figure 1 represents the distributions for first two PCs with respect to target class, original attributes and corresponding impacts of the target classes within the Dataset 2 and Dataset 3. These plots also indicate the non-linearity of the problem specifically in terms of first two PCs covering the highest variances within the overall principal components. However, the plots help to understand the corresponding influence of the variables within the datasets on the classification of phishing and legitimate websites. For instance, in Fig. 1(A), *'web-traffic'* has a clear impact on class '1' while *'ssl-final-* *state'* influences the '-1' class. The first two PCs cover approximately 42% of the overall PCs variance.

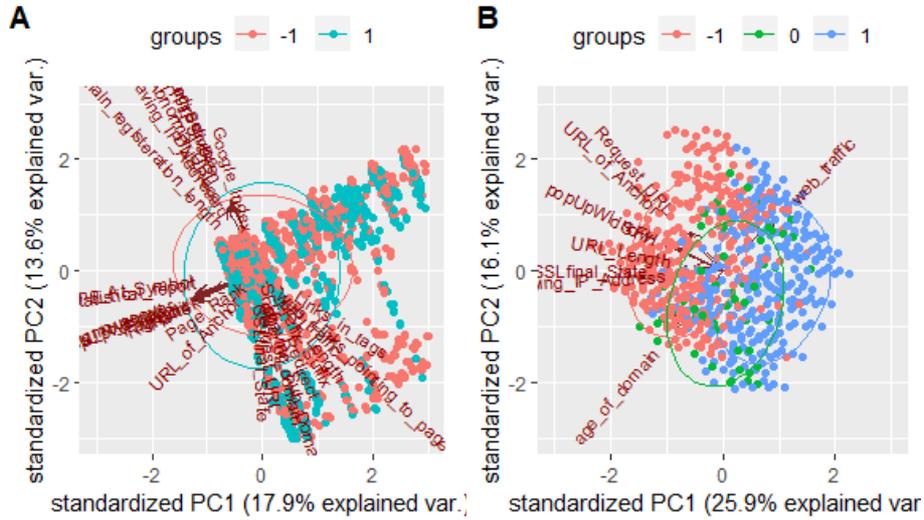

**Fig. 1(A).** First two PCA components' distributions in Dataset 2; **(B).** First two PCA components' distributions in Dataset 3

The correlation coefficient between the dataset attributes is represented by the principal components' loadings (i.e. obtained through PCA). The component rotations provide the maximized sum of variances of the squared loadings. The absolute sum of component rotations gives the degree of importance for the corresponding attributes in dataset.

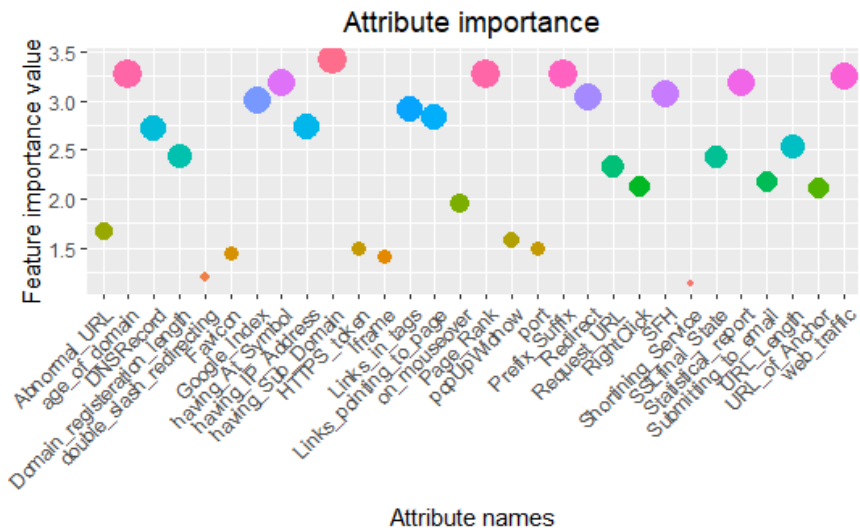

**Fig. 2.** Measure of feature importance within the Dataset 2 using PCA

Fig. 2 demonstrates the attribute/feature importance in Dataset 2 which is calculated through the PCs loadings. The result indicates a clear variation in the importance measure of variables that might be helpful to eliminate the unnecessary features from the dataset. For instance, *'having-sub-domain'* and *'age-of-domain'* are indicated the top-ranked variables compared to *'double-slash-redirecting'* and *'shortening-service'* which are indicated the least important variables within the Dataset 2.

## 4 Results and Discussion

Results are achieved following the experimental design for two experiments (A and B) and the variables rankings in three Datasets. Table 1, Table 2 and Table 3 presents the statistical results achieved by different ML algorithms over the three datasets without dimensionality reductions.

### 4.1 Classification Performance Using Original Datasets and ML Algorithms

Table 1 presents the classifiers outcomes for dataset 1. The highest accuracy is achieved by RF and ANN indicating 97.87% and 97.83% respectively while NB indicated the lowest accuracy (82.17%). It can be observed that the NB classifier has very high specificity (96.49%) however, very low recall (68.2%) which indicate poor compromise between the sensitivity and specificity from NB, therefore affecting the overall accuracy of the classifier.

**Table 1.** Classifier Performance Over Dataset 1 Using All Features

| Classifier | Accuracy | Specificity | Prcision | Recall | F1 Score |
|---|---|---|---|---|---|
| D- Tree | 95.73% | 95.48% | 95.61% | 95.98% | 95.80% |
| SVM | 94.37% | 93.59% | 93.83% | 95.13% | 94.48% |
| RF | 97.87% | 98.45% | 98.47% | 97.30% | 97.88% |
| NB | 82.17% | 96.49% | 95.22% | 68.20% | 79.48% |
| KNN | 94.00% | 95.68% | 95.64% | 92.36% | 93.97% |
| ANN | 97.83% | 97.91% | 97.96% | 97.76% | 97.86% |

**Table 2.** Classifier Performance on Dataset 2 Using All Features

| Classifier | Accuracy | Specificity | Prcision | Recall | F1 Score |
|---|---|---|---|---|---|
| D- Tree | 95.30% | 94.65% | 95.71% | 95.82% | 95.77% |
| SVM | 92.58% | 90.40% | 92.22% | 94.62% | 93.40% |
| RF | 95.96% | 94.51% | 95.67% | 97.12% | 96.39% |
| NB | 60.48% | 99.80% | 99.44% | 28.95% | 44.85% |
| KNN | 93.10% | 92.21% | 93.76% | 93.81% | 93.78% |
| ANN | 95.90% | 95.93% | 96.71% | 95.87% | 96.29% |

Table 2 indicates the supremacy of RF and ANNs in terms of classifying the phishing websites for Dataset 2 while using the original attributes. Keeping in mind that dataset 2 has 30 different features and contains only 11000 distinct entries, still our employed ML models are achieving state of the art performance. In study [12], the authors achieved an accuracy of 95% using ANNs, however that study was focusing the classification of phishing emails and not websites. In the current work for classification of phishing websites, the ANN performed slightly higher by achieving accuracy of 97% for Dataset 1 and 96% for Dataset 2.

**Table 3.** Classifier Performance On Dataset 3 Using All Features

| Classifier | Accuracy | Specificity | Prcision | Recall | F1 Score |
|---|---|---|---|---|---|
| D-Tree | 91.63% | 93.72% | 87.44% | 87.44% | 87.44% |
| SVM | 89.66% | 92.24% | 84.48% | 84.48% | 84.48% |
| RF | 92.94% | 94.70% | 89.41% | 89.41% | 89.41% |
| NB | 89.33% | 92.0% | 83.99% | 83.99% | 83.99% |
| KNN | 91.30% | 93.47% | 86.95% | 86.95% | 86.95% |
| ANN | 90.48% | 92.86% | 85.71% | 85.71% | 85.71% |

The results achieved by the employed classification algorithms on Dataset 3 are shown in Table 3. It can be observed that the outcomes from all ML algorithms are slightly lower than the performances in case of Dataset 1 and Dataset 2. The highest accuracy obtained by any classifier on Dataset 3 is 92.94% which is (4%) lower than what is observed in case of Dataset 1 and Dataset 2. However, there are some factors that should be noted in this case. Primarily, the Dataset 3 is small (i.e. only 1353 distinct instances) with small number of distinguishing features (9 attributes) as compared to more than 10000 instances for Datasets 1 and Dataset 2. Likewise, feature set in later case are 48 (in Dataset 1) and 30 (in Dataset 2). Furthermore, Dataset 3 is multi-class problem (3-classes) as compared to bi-class problem in case of Dataset 1 and 2. This makes the classification task more challenging specifically when training data is limited as well. These factors influence the overall accuracy of our ML models and hence indicating relatively low performance in this case which was expected. More specifically, for a multiclass dataset, there needs to be sufficiently large number of distinct instances which the classifier can study and then try to make predictions. If a dataset is small as well as multiclass, the performances are expected to be mediocre.

From statistical results presented in Table 1-3, it can be seen that overall performance from all ML algorithms are quite satisfactory except NB which performed relatively low for the Dataset 1 and 2, however indicated relatively better in case of Dataset 3. One interesting aspect of the current study is the use of Dataset 3 for phishing website classification. This dataset was not used by existing studies to the best of authors' knowledge. This may be due to the limited size of this dataset however, it might be helpful to investigate the classification performance using this dataset as it defines different attributes/features to other datasets. Furthermore, this dataset is multi-class as compared to Dataset 1 and 2 which are bi-class dataset, hence, this study helps in getting even more better insights, as it is the only publicly available multiclass phishing websites dataset.

### 4.2   Classification Performance After Dimensionality Reduction using PCA

Table 4, Table 5 and Table 6 presents the statistical results achieved by different ML algorithms for the PCA-based dimension reduced datasets.

**Table 4.** Classifier Performance On Dataset 1 After PCA

| Classifier | Accuracy | Specificity | Prcision | Recall | F1 Score |
|---|---|---|---|---|---|
| D-Tree | 91.83% | 91.29% | 91.58% | 92.00% | 91.97% |
| SVM | 93.97% | 93.05% | 93.33% | 94.87% | 94.09% |
| RF | 94.90% | 96.49% | 96.46% | 93.35% | 94.88% |
| NB | 78.37% | 89.13% | 86.49% | 67.87% | 76.06% |
| KNN | 93.97% | 95.61% | 95.57% | 93.36% | 93.94% |
| ANN | 97.13% | 96.22% | 96.48% | 98.03% | 97.19% |

It can be seen that ANN outperformed other classifiers when trained and tested over the PCA based dimension reduced Dataset 1. An accuracy of 97.13% is achieved while using first 30 components which contain the 95% variance of the overall PCs distributions. Dimensionality reduction is not previously employed in this specific domain of classification of phishing websites, this work is first of its kind and it is indeed getting good results while using reduced data and ANN in case of Dataset 1.

**Table 5.** Classifier Performance On Dataset 2 After PCA

| Classifier | Accuracy | Specificity | Prcision | Recall | F1 Score |
|---|---|---|---|---|---|
| D- Tree | 92.58% | 91.12% | 92.95% | 93.75% | 93.35% |
| SVM | 92.43% | 89.57% | 91.89% | 94.73% | 93.29% |
| RF | 93.79% | 92.82% | 94.26% | 94.57% | 94.41% |
| NB | 90.50% | 85.43% | 89.01% | 94.57% | 91.70% |
| KNN | 92.85% | 91.80% | 93.45% | 93.70% | 93.57% |
| ANN | 94.33% | 95.46% | 96.25% | 93.43% | 94.82% |

Similarly, first 18 PCs covers the 95% of overall components variance for the Dataset 2 which originally consists of 30. Table 5 indicates that the classification performances are relatively lower than the Table 4 for the Dataset 1 however, it is expected because the dataset 2 comprises of comparatively less features than dataset 1. The overall performance is satisfactory though more specifically, we can see the balance between the sensitivity and specificity. This factor is very interesting because it validates the best compromise between true and false positives from a classifier.

**Table 6.** Classifier Performance On Dataset 3 After PCA

| Classifier | Accuracy | Specificity | Prcision | Recall | F1 Score |
|---|---|---|---|---|---|
| D- Tree | 90.31% | 92.73% | 85.47% | 85.47% | 85.47% |
| SVM | 89.16% | 91.87% | 83.74% | 83.74% | 83.74% |
| RF | 90.15% | 92.61% | 85.22% | 85.22% | 85.22% |
| NB | 89.00% | 91.75% | 83.50% | 83.50% | 83.50% |
| KNN | 92.12% | 94.09% | 88.18% | 88.18% | 88.18% |
| ANN | 91.13% | 93.35% | 86.70% | 86.70% | 86.70% |

Table 6 shows the summary of statistical results performed by the aforementioned classifiers while trained and tested over the PCA based dimension reduced Dataset 3. Similar to previous results, the outcomes indicated PCA to be handy on Dataset 3 as well. By training on data produced by PCA, K-neighbors and ANN, performed even better than training on the whole datasets having all the features.

**Table 7.** Top-ranked Features Identified within Three Datasets using PCA

| Feature Rank | Features from Dataset 1 | Features from Dataset 2 | Features from Dataset 3 |
|---|---|---|---|
| 1 | RandomString | having_Sub_Domain | Request_URL |
| 2 | DomainInPaths | age_of_domain | popUpWidnow |
| 3 | NumUnderscore | Page_Rank | URL_of_Anchor |
| 4 | RightClickDisabled | Prefix_Suffix | SSLfinal_State |
| 5 | ExtFavicon | web_traffic | URL_Length |
| 6 | NumPercent | Statistical_report | having_IP_Address |
| 7 | NumSensitiveWords | having_At_Symbol | SFH |
| 8 | EmbeddedBrandName | SFH | web_traffic |
| 9 | TildeSymbol | Redirect | age_of_domain |
| 10 | SubmitInfoToEmail | Google_Index | |

Table 7 shows the top 10 ranked features within the three datasets identified by the PCA based on attribute loadings in components as described earlier (Section 3.3). It can be observed in Table 7 as well as Fig. 2 that the most important features in Dataset 2 for instance, are the *'having-sub-dmian'* and *'age-of-domain'* while *'request-url'* and *'popUpWindow'* in Dataset 3. The investigation of such distinguishing features would be helpful for domain experts and research community in this domain to further explore the varying combinations of only top-ranked features within the various datasets that might be helpful for further optimization of cyber security applications.

## 5  Conclusion and Future Work

This manuscript aims a comprehensive analysis of various ML algorithms to classify the fishing websites using multiple datasets. The study investigated the ranking of attributes within different datasets while utilizing the well-known dimensionality reduction technique (i.e. PCA). We further conducted experiments on various datasets with and without dimension reductions using PCA and compared the performances of the state-of-the-art ML algorithms. The statistical results indicated the vital role of PCA specifically for eliminating the irrelevant features from the original datasets while not affecting the classification accuracy. Furthermore, this study is first of its kind to investigate the top-ranked attributes (features) within different datasets which might be useful for furthering the research within domain of cyber security. For instance, it would be helpful to investigate the formation of a dataset consisting the integrated features identified in this study and then use the ML techniques to classify the more complex problems (i.e. adversarial attacks) in this domain. Likewise, the existing datasets such as Dataset 3 can further be extended that might be helpful to improve the classification performance.